\documentclass[12pt,english]{article}
\usepackage[T1]{fontenc}
\usepackage[latin1]{inputenc}
\usepackage{graphics}
\usepackage{babel}
\begin{document}

\begin{center}
{\Large Exact Classical Correspondence in Quantum Cosmology}
 
\vspace{.5 in}

Moncy V. John

{\it Department of Physics, St. Thomas College, Kozhencherri - 689641, Kerala, India

moncyjohn@yahoo.co.uk} 

\end{center} 
\vspace{.5in}

\abstract{We find a Friedmann model with appropriate matter/energy density  such  that the  solution of the Wheeler-DeWitt equation exactly corresponds to the classical evolution. The well-known problems in quantum cosmology disappear in the resulting coasting evolution. The exact quantum-classical correspondence is demonstrated with the help of  the de Broglie-Bohm and modified de Broglie-Bohm approaches to quantum mechanics. It is reassuring that such a solution leads to a robust model for the universe, which  agrees well with cosmological expansion indicated by SNe Ia data.}

\section{Introduction}

Just as  quantum mechanics  contains   classical mechanics as a limiting case,  a  quantum gravity theory must have the  general theory of relativity as its classical limit. In line with these, it is natural to require that  the classical limit of quantum cosmology (which is the  quantum theory applicable to the universe)    must be a robust cosmological model capable of describing the state of the universe during its present epoch. In quantum cosmology, the universe is taken to be described by a  wave function,  which obeys the Wheeler-DeWitt (WD) equation \cite{wd}. However, there are some deeper conceptual problems in this theory when compared to standard quantum mechanics. One such problem is related to the use of probability in this scheme. The WD equation does not contain the time parameter and is written  as $\hat{H}\Psi =0$. In the much simplified version of a mini-superspace model in quantum cosmology, this equation is similar to a zero energy Schrodinger equation and  leads to a stationary wave function independent of time. Since there are no excited states in this case, the superposition of energy eigenfunctions cannot be done.  This makes the application of the superposition principle, and consequently that of the probability axiom, meaningless. Another problem is  that the role of the observer in this scheme is not clear.  It suffers a  basic  difficulty that the observer of this quantum system is  part of the  system itself, which is the universe, and is not just an outside `classical' observer. Consequently there cannot be any 'wave function collapse' as in  standard quantum mechanics. It is hoped that a proper resolution of these problems can give   impetus to a deeper understanding of cosmology. 

We restrict ourselves to the mini-superspace model in quantum cosmology. The  work proposes to  identify a Friedmann model with an appropriate energy/matter content in the WD equation, such that it has exact classical correspondence.  The idea of exact classical correspondence is introduced in the sense that the  WD equation reduces to the classical Hamilton-Jacobi-type equation in cosmology under the substitution $\Psi=\exp{i\hat{S}/\hbar}$.  We find that an energy/matter density in the universe with a particular  equation of state parameter makes the WD equation equivalent to that of the  classical equation. Obviously, since in this scenario the universe is always classical, there is no need to invoke the probaility axiom. Similarly, there is no need of wave function collapse, etc., for the universe. Most important is the fact  that the resulting cosmological model has very good predictions for the present  universe.

The equivalence of classical and quantum evolution is more rigorously shown by making use of  the de Broglie-Bohm (dBB) \cite{dBB,bohm,holland,bohmhiley} and the modified  de Broglie-Bohm (MdBB) \cite{mvj1} approaches to quantum mechanics. Recently, there is a renewed interest in quantum trajectories in real space, such as the dBB trajectories and those due to  Floyd, Faraggi and Matone (FFM) \cite{carroll,pratim,wyatt2}.  These real trajectory representations could provide sensible, ontological interpretation to several quantum phenomena. MdBB was the result of a similar  attempt, where complex quantum trajectories were first conceived and drawn by modifying the dBB  approach. Such trajectories were drawn  in  \cite{mvj1} for the cases of the harmonic oscillator, the particle approaching a potential step, a free particle, a time-dependent spreading wave packet etc.  It was found that MdBB has several advantages over the conventional dBB trajectories, where problems such as stationarity of particles in bound eigenstates etc., exist \cite{holland}. Our attempt in  this paper would be to apply the methods of both dBB and MdBB  quantum trajectories to quantum cosmology and to study the  quantum-classical correspondence for a   class of   models, which procedure helps us to  identify the cosmological model having the desired exact correspondence. 

We note that in quantum cosmology, obtaining the evolution of the scale factor $a(t)$ using the dBB approach is already familiar. A recent work \cite{he_gao_cai}, which uses this method, shows that the universe can spontaneously and irreversibly be created from nothing, and also that  it can then pass over to an inflationary epoch. The authors first obtain the solutions to the WD equation for an empty  universe, with closed, open or flat space sections, when a particular operator ordering is chosen in it. The evolutions of the scale factor $a(t)$ for the early universe, corresponding to these solutions, are found using the dBB formalism. The present work too follows a similar approach. The prospects of relating one of these  early evolutions to the late classical `coasting' evolution obtained in the present work seems to be an interesting open problem and we discuss it in the last section.

The paper is organised in the following way. We review in section 2  the basics  of the dBB and the MdBB  trajectory representations in quantum mechanics. In Sec. 3,  the theoretical framework of the Friedmann models in classical cosmology is outlined. The Wheeler-De Witt (WD) equation in quantum cosmology is introduced in the next section. In Sec. 5, using the dBB and MdBB approaches, we analyze the behaviour of  a class of Friedmann models during the quantum to classical transition. The results are discussed in the last section.

\section{Quantum trajectories}
\subsection{Classical mechanics and analogy with wave optics}
A well-known formulation of  classical  mechanics of a system of particles, other than the Lagrangian and Hamiltonian formulations, is based on solving the Hamilton-Jacobi equation

\begin{equation}
\frac{\partial S}{\partial t}+ H\left( q_i,\frac{\partial S}{\partial q_i },t\right) =0, \label{chje}
\end{equation}
where $S(q_i, \alpha_i,t)$ is  the Hamilton's principal function, $H$ is the Hamiltonian,  $q_i$ are the configuration space variables, $\alpha_i$ are  constants of integration and $t$ denotes time. If the Hamiltonian  is independent of $t$ and is a constant of motion equal to the total energy $E$, then $S$ and the Hamilton's characteristic function $W$ are related by

\begin{equation}
S=W-Et. \label{eq:s}
\end{equation}
Since $W$ is independent of time, the surfaces with $W$ = constant have fixed locations in configuration space. However, 
 the  $S$ = constant  surfaces move in time and may be considered as wave fronts propagating in this space \cite{goldstein}. For a single particle one can show that the wave velocity at any point is given by

\begin{equation}
u=\frac{E}{\mid \nabla W \mid} = \frac{E}{p}=\frac{E}{mv}. \label{u}
\end{equation}
This shows that the velocity of a point on this surface is inversely proportional to the spatial velocity of the particle. Also one can show that the trajectories of the particle must always be normal to the surfaces of constant $S$. The momentum of the particle is obtained as \cite{goldstein}

\begin{equation}
{\bf p} =\nabla W.
\end{equation}
Thus the particle trajectories orthogonal to surfaces of constant $S$ in classical mechanics, as we have discussed above, are similar to light rays traveling orthogonal to Huygens' wavefronts. It was this analogy with wave optics that de broglie utilised to formulate his principle of wave-particle duality. 

\subsection{de Broglie-Bohm quantum trajectory formulation}

The fact that the 1924 Ph.D. thesis of de Broglie \cite{dbthesis} contained the seed of a new mechanics, which can replace classical mechanics, did not get due attention for a long time. In it, and later in a paper published in 1927 \cite{dBB}, he proposed  that Newton's first law of motion be abandoned, and replaced by a new postulate, according to which a freely moving body follows a trajectory that is orthogonal to the surfaces of equal phase of an associated guiding wave. He  presented his new theory in the 1927 Solvay conference too, but it was not well-received at that time \cite{valentini}.

The Schrodinger equation, which is the corner stone of  quantum mechanics, can be recast in a form remniscent of Hamilton-Jacobi equation  by substituting $\Psi =R\exp (iS/\hbar )$ in it. This results in two equations. The first  is

\begin{equation}
\frac {\partial {S}}{\partial t} + \left[ \frac{1}{2m}\left( 
{\nabla  {S}}\right)^2 +V\right]  =
\frac{\hbar^2}{2m} \frac{\nabla^2 {R}}{R}, \label{eq:dbb_qhje}
 \end{equation}
which resembles the classical Hamilton-Jacobi equation. The second is

\begin{equation}
\frac{\partial \rho}{\partial t}+\nabla . \left( \rho \frac{\nabla S}{m}\right) =0,
\end{equation}
which is the standard continuity equation  obeyed by the Born's probability density $ \rho \equiv \Psi^{\star}\Psi = R^2$. The  guidance relation 
for particle trajectories, as proposed by de Broglie, is

\begin{equation}
m_i \dot{{\bf x}}_i=\nabla _i S. \label{eq:dbeqnmtn}
\end{equation}
Solving this, we get a particle trajectory for each initial position.
In 1952, David Bohm rediscovered this formalism \cite{bohm} in a slightly different way. He noted that if we take the time dervative of the above equation,  the Newton's law of motion may be obtained in the modified  form

\begin{equation}
m\ddot{\bf x}_i=-\nabla _i(V+Q). \label{eq:beqnmtn}
\end{equation}
Here

\begin{equation}
Q=-\Sigma \frac{\hbar ^2}{2m_i} \frac{\nabla^2_i \mid \Psi \mid }{\mid \Psi \mid} \label{eq:qmpot}
\end{equation}
 is called the `quantum potential'. This results in the same mechanics as that of de Broglie, but in the language of Newton's equation with the above unnatural quantum potential. We shall note that the first order equation of motion (\ref{eq:dbeqnmtn}) suggested by de Broglie represented a unification of the principles of mechanics and optics and this should be distinguished from the Bohm's second order dynamics based on (\ref{eq:beqnmtn}) and (\ref{eq:qmpot}) \cite{valentini}.  Bohm's revival of de Broglie's theory in the  pseudo-Newtonian form has led to a mistaken notion that de Broglie-Bohm (dBB) theory constituted a return to classical mechanics. In fact, de Broglie's theory was a new formulation of dynamics in terms of wave-particle duality.

It is well-known that neither the particle picture nor the trajectories have any role in the  standard Copenhagen interpretation of quantum mechanics and that several physicists, including Albert Einstein, were not happy with the Copenhagen interpretation for various reasons.   Einstein was not happy with the de Broglian mechanics either \cite{valentini}. The reason for the latter was speculated to be the problem it faced with real wave functions where the phase gradient  $\nabla S =0$. For the wide class of bound state problems, the time-independent part of the wave function is real and hence   while applying the equation of motion (\ref{eq:dbeqnmtn}),     the velocity of the particle turns out to be zero everywhere. This feature, that the particles in bound eigenstates are at rest irrespective of their position and energy, is counter-intuitive and is not a satisfactory one.

\subsection{Complex quantum trajectories}
 The modified dBB approach  also  attempts to unify the principles of mechanics and optics, but in a different way. An immediate    application of the formalism was to solve the above problem of stationarity of quantum particles in bound states, in a natural way \cite{mvj1}. For obtaining this representation, we shall substitute $\Psi=e^{i\hat{S}/\hbar}$ in the Schrodinger equation  to obtain a single equation, which is now called the quantum Hamilton-Jacobi equation (QHJE) \cite{goldstein}

\begin{equation}
\frac {\partial \hat{S}}{\partial t} + \left[ \frac{1}{2m}\left( \frac
{\partial  \hat{S}}{\partial x}\right)^2 +V\right]  =
\frac{i\hbar}{2m} \frac{\partial^2 \hat{S}}{\partial x^2}. \label{eq:qhje}
 \end{equation}
In this section, we restrict ourselves to one dimension. An  equation of motion for the particle, similar to that used by de Broglie, can now be defined:

\begin{equation}
m\dot{x} \equiv \frac {\partial \hat{S}}{\partial x}= \frac {\hbar
}{i} \frac {1}{\Psi}\frac {\partial \Psi}{\partial x}.
\label{eq:xdot}   \end{equation}
  The trajectories $x(t)$   are obtained by integrating this equation with respect to time, with various initial positions. In general,  they  lie in a complex $x$-plane where $x\equiv x_r + ix_i$. We find that even real wave functions produce non-stationary trajectories. The reason for this is  that the  identification $\Psi=e^{i\hat{S}/\hbar}$  helps to utilize all the information contained in $\Psi$ for obtaining the trajectory. (In contrast, the dBB approach, which uses  $\Psi=R e^{i{S}/\hbar}$, does not have this advantage.) The observable trajectory of the particle may be considered to be the real part $x_r(t)$ of the complex trajectories \cite{mvj1}.

 When $\hat{S}=\hat{W}-Et$, where  $E$ and   $t$ are assumed to be real, the Schrodinger equation gives us an expression for the energy of the particle

 \begin{equation}
 E=\frac{1}{2} m\dot{x}^2+V(x)+\frac{\hbar}{2i} \frac{\partial
  \dot{x}}{\partial x}. \label{eq:E}
   \end{equation}
  The last term resembles the quantum potential in the dBB theory. However, the concept of quantum potential is not an integral part of this formalism since the equation of motion (\ref{eq:xdot})  adopted here  is not based on it.

The complex eigentrajectories in the free particle, harmonic oscillator and potential step problems and complex trajectories for a wave packet solution were obtained in \cite{mvj1}. As an example, complex trajectories in the first excited state of harmonic oscillator is shown in figures 1.

\begin{figure}[ht] 
\centering{\resizebox {0.5 \textwidth} {0.3 \textheight }  
{\includegraphics {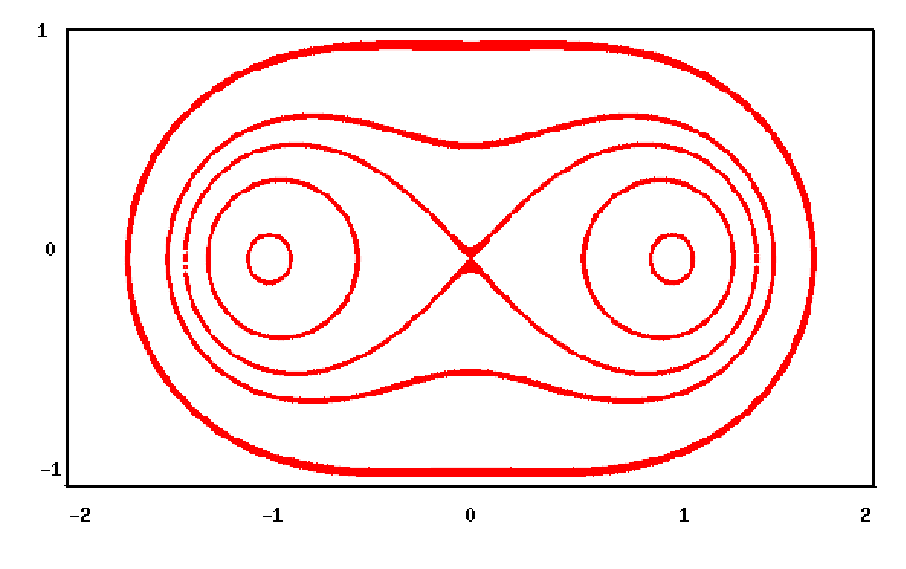}} 
\caption{ The complex trajectories for the first excited state of harmonic oscillator.   }  \label{fig:n1shmtraj}} 
   \end{figure}

 This  formalism was  extended to three dimensional problems such as the hydrogen atom  by Yang \cite{yang} and  was used to investigate one dimensional scattering problems and bound state problems by  Chou and Wyatt \cite{wyatt}. Later, a complex trajectory approach for solving the QHJE was developed  by Tannor and
co-workers \cite{goldfarb}.  Sanz and Miret-Artes  found the complex trajectory representation  useful in  better understanding the  nonlocality in quantum mechanics \cite{sanz}. The role of probability in this scheme was explained and extension of the probability density to the complex plane was performed in \cite{mvj2}. Certain characteristics such as identical classical and MdBB quantum trajectories for particles in coherent states were pointed out in \cite{mvj3}.

\section{Classical  cosmology}

\subsection {Field equations as Euler-Lagrange equations}

The Einstein equations in cosmology can be obtained as Euler-Lagrange equations using an action principle. For the maximally spatially symmetric spacetime described by a Robertson-Walker (RW) metric \cite{cosmo_cl_book}

\begin{equation}
ds^{2} = c^2dt^{2} - a^{2}(t) \left[ \frac {dr^{2}}{1-kr^{2}} +
r^{2} (d\theta ^{2} + \sin ^{2}\theta\; d\phi ^{2})\right] ,
\label{eq:rwle} 
\end{equation}
 under the ADM 3+1 split \cite{kolbturner_book}, the action is 

\begin{eqnarray}
I & = & 2\pi ^{2}c^2\int Na^{3}\left[ -\frac {1}{16\pi G}\left( \frac
{6}{N^{2}}\frac {\dot {a}^{2}}{a^{2}} - \frac {6kc^2}{a^{2}} \right) +
 \frac {\dot {\phi
}^{2}}{2N^{2}} - V(\phi )\right] dt  \nonumber \\
& \equiv & \int L\; dt. \label{eq:ielphi}
\end{eqnarray}
This is when the universe contains only a spatially homogeneous scalar field $\phi$.
Here $N$ is called the lapse function, for which one can fix some convenient gauge. Using the Lagrangian $L$, we may write the Euler-Lagrange equations
for the variables
$N$ and $\phi $   and obtain the   Einstein equations in this case as

\begin{equation}
\frac {\dot {a}^{2}}{a^{2}} + \frac {kc^2}{a^{2}} = 
 \frac {8\pi G}{3c^2} \left[ \frac {\dot {\phi }^{2}}{2} + V(\phi
)\right] ,\label{eq:t-tphi}
\end{equation}
and

\begin{equation}
\ddot {\phi }+ 3 \frac {\dot {a}}{a}
\dot {\phi } + \frac {dV(\phi ) }{d\phi }=0, \label{eq:consphi}
\end{equation}
by fixing the gauge $N=1$.

Similarly, for a de Sitter model which contains only a cosmological
constant, the Lagrangian can be taken to be

\begin{equation}
L=2\pi ^{2} Na^{3}c^2\left[ -\frac {1}{16\pi G}\left( \frac
{6}{N^{2}}\frac {\dot {a}^{2}}{a^{2}}- \frac {6kc^2}{a^{2}} \right) -
 \rho _{\lambda }\right] \label{eq:iellambda}
\end{equation}
The Einstein equations

\begin{equation}
\frac {\dot {a}^{2}}{a^{2}} + \frac {kc^2}{a^{2}} 
 = \frac {8\pi G}{3} \rho _{\lambda }, \label{eq:t-tdS}
\end{equation}

\begin{equation}
2\frac {\ddot {a}}{a} + \frac {\dot {a}^{2}}{a^{2}} + \frac
{kc^2}{a^{2}}  = 8\pi G \rho _{\lambda }. \label{eq:s-sdS}
\end{equation}
 are
obtained on writing the Euler-Lagrange equations corresponding to
variations with respect to $N$ and $a$, in the gauge $N=1$.

\subsection{Hamiltonian formulation}

 As in the above, we consider a 
model in which the only degrees of freedom  are those of the scale
factor $a$ of a  RW spacetime and a spatially homogeneous
scalar field $\phi $. The Lagrangian for this problem is given by
(\ref{eq:ielphi}) as 

\begin{equation}
L=-\frac {3\pi }{4G}Nc^2\left[ \frac {a\dot {a}^{2}}{N^{2}}-kc^2 a-\frac
{8\pi G}{3}\left( \frac {\dot {\phi }^{2}}{2N^{2}}-V(\phi )\right)
a^{3}\right], 
\end{equation}

From this we can find the conjugate momenta $\pi _{a}$ and $\pi _{\phi
}$ to $a$ and $\phi$ as

\begin{equation}
\pi _{a}=\frac {\partial L}{\partial \dot {a}} = -\frac {3\pi }{2G}c^2
\frac {a \dot {a}}{N} \label{eq:pia}
\end{equation} 
and 

\begin{equation}
\pi _{\phi }= \frac {\partial L}{\partial \dot {\phi }} = 2\pi
^{2}c^2\frac {a^{3}\dot {\phi }}{N}.
\end{equation} 
The canonical Hamiltonian can now be constructed as 

\begin{eqnarray}
{\cal H}_{c}& = &\pi _{a}\dot {a}+ \pi _{\phi }\dot {\phi } -L \nonumber \\
&=& N\left[ -\frac {G}{3\pi c^2}\frac {\pi _{a}^{2}}{a}-\frac {3\pi
}{4G} kc^4 a+\frac {3\pi }{4G}a^{3}\left( \frac {G}{3\pi ^{3}c^2}\frac {\pi _{\phi
}^{2}}{a^{6}} +\frac {8\pi G}{3} V(\phi )\right) \right] \nonumber \\
&\equiv & N{\cal H}.
\end{eqnarray} 
The secondary constraint now give 

\begin{equation}
{\cal H}=  -\frac {G}{3\pi c^2}\frac {\pi _{a}^{2}}{a}-\frac {3\pi
}{4G} kc^4 a+\frac {3\pi }{4G}a^{3}\left( \frac {G}{3\pi ^{3}c^2}\frac {\pi _{\phi
}^{2}}{a^{6}} +\frac {8\pi G}{3} V(\phi )\right) =0, \label{eq:hamphi}
\end{equation} 
which is equivalent to (\ref{eq:t-tphi}). For the RW spacetime which
contains only a cosmological constant, (\ref{eq:iellambda}) helps us
to write the constraint equation  as

\begin{equation}
{\cal H}=  -\frac {G}{3\pi c^2}\frac {\pi _{a}^{2}}{a}-\frac {3\pi
}{4G} kc^4 a+\frac {3\pi }{4G}a^{3} 
\frac {8\pi G}{3} \rho _{\lambda } =0. \label{eq:haml}
\end{equation} 

This equation is equivalent to (\ref{eq:t-tdS}). 
The fact that ${\cal H}=0$ is a consequence of a new symmetry
of the theory, namely, time reparametrisation invariance. This means
that using a new time variable $t^{\prime }$ such that $dt^{\prime }=
N\;dt$ will not affect the equations of motion. Also this enables one
to choose some convenient gauge for $N$.

\section{Wheeler-DeWitt equation in quantum cosmology}
\label{sec-wd}

Canonical quantisation of  a classical system like the one above
means introduction of a wave function $\Psi (a, \phi )$
and requiring that it satisfies \cite{kolbturner_book}

\begin{equation}
i\hbar \frac {\partial \Psi }{\partial t} = {\cal H}_{c}\Psi = N{\cal H}\Psi. \label{eq:gwd}
\end{equation}
To ensure that time reparametrisation invariance is not lost at the
quantum level, the conventional practice is to ask that the wave
function is annihilated by the operator version of ${\cal H}$; i.e.,

\begin{equation}
{\cal H}\Psi =0. \label{eq:wd}
\end{equation} 
Eq. 
(\ref{eq:wd}) is called the Wheeler-DeWitt (WD) equation. 

This equation is analogous to a zero energy Schrodinger equation, in
which the dynamical variables $a,\; \phi $ etc. and their
conjugate momenta $\pi _a,\; \pi _{\phi }$ etc. 
 are replaced by the corresponding operators. The wave
function $\Psi $ is defined on the minisuperspace with just one coordinate $a$ and we expect it to
provide information regarding the evolution of the universe. An
intriguing fact here is that the wave function is independent of
time; they are stationary solutions in the minisuperspace. 

The transition from the quantum cosmology era, which is the earliest epoch after the big bang, to the late classical era is expected to happen when the scale factor $a$ exceeds the Planck length $l_p=\sqrt{\hbar G/c^3}$, which is a quantity with dimension of length, constructed using the three fundamental constants. 
The usual approach in quantum cosmology to study this transition  is to check whether the wave function of the universe $\Psi$ is strongly peaked around the trajectories $a(t)$ identified by the classical solutions \cite{halli}.
The wave functions commonly arising in quantum cosmology are of WKB
form and may be broadly classified as oscillatory, of the form
$e^{iS/\hbar}$ or exponential, of the form $e^{-I/\hbar}$.
The wave function of the form $e^{iS/\hbar}$ is normally thought of as
being peaked about a set of solutions to the classical equations and
hence predicts classical behaviour. A wave function of the form
$e^{-I/\hbar}$ predicts no correlation between coordinates and momenta and
so cannot correspond to classical behaviour. If  Born's probability axiom is assumed to be valid  in the case of the universe, one can argue that  $\Psi $ must be strongly peaked around the  trajectories identified by the classical solutions.

\section{Classical correspondence in quantum cosmology}

The  application of Born's probability is criticised  on the grounds that the role of the observer and the observed is not clear in this case. As mentioned in the Introduction, the observer is part of the observed system, which is the universe. This is unlike  the situation in standard   quantum mechanics, where the quantum system is always observed by an outside classical observer. Moreover, since we have only the zero-energy  ground state solution for the WD equation (\ref{eq:wd}),  superposition principle cannot work  and therefore the application of probability axiom is meaningless.  

Instead, in this paper we approach the problem of classical correspondence by rewriting the WD equation as the  quantum Hamilton-Jacobi equation in cosmology, and then analysing  the quantum trajectories obtained from it. The quantum cosmological  Hamilton-Jacobi equation (QCHJE) (which is akin to QHJE  obtained from the Schrodinger equation),  can be derived by substituting $\Psi = e^{i\hat{S}/\hbar}$ in the WD equation. 
Instead, if we can identify $\hat{S}$ from the solution of the WD equation, quantum trajectories can be found by integrating a de Broglie-type equation of motion. 
If  the classical and quantum trajectories are found to be identical, there is exact classical correspondence. We do not expect this  to happen for all  potentials. 
 The above is an acceptable  criterion for exact classical correspondence since what we try to find is whether  the WD  equation in this case tends to the corresponding classical equation  and is devoid of $\hbar$.

It is  instructive to take the free particle ($V=0$) case in ordinary quantum theory as an example. Substituting $\psi =  e ^{i\hat{S}/\hbar}$ into the Schrodinger equation, one gets the QHJE  as (See Sec. 2.3)

\begin{equation}
\left[ \frac{1}{2m} (\nabla \hat{S})^2 +V\right] +\frac{\partial \hat{S}}{\partial t} = \frac{i\hbar}{2m}\nabla^2 \hat{S}.\label{eq:qhje_3d}
\end{equation} 
This would be the classical Hamilton-Jacobi (HJ) equation for the Hamilton's principal function $\hat{S}$, if   the  right hand side (the quantum potential)  vanishes. A more rigorous approach to obtain exact quantum-classical correspondence would be to  see whether the quantum trajectories obtained in a particular problem are the same as its classical trajectories.  One can  see that in the above case in ordinary quantum mechanics, plane waves in the dBB/MdBB schemes  result  in  rectilinear motion, which is the free particle motion  in classical mechanics.  In other words, the plane wave solution can be used to describe free particles as if they are classical particles and this  justifies the use of plane waves to describe incident particles in quantum scattering problems. The dBB scheme predicts identical classical and quantum trajectories only in this zero potential region. MdBB shows such property for the ground state harmonic oscillator and for several coherent states \cite{mvj3,fring}.  

 In quantum cosmology, for finding a potential that  results in exact classical correspondence, we note that the Lagrangian for the Friedmann model  of the universe is of the form 

\begin{equation}
L=2\pi^2a^3Nc^2 \left[ - \frac{1}{16\pi G}\left( \frac{6}{N^2}\frac{\dot{a}^2}{a^2} - \frac{6kc^2}{a^2}\right) -\frac{C_n}{a^n} \right]. \label{eq:L}
\end{equation}
In the Friedmann model, the total density of the universe varies as $\rho \propto \frac{1}{a^n}$. The index $n$ is related to the equation of state parameter $w$ (that appears in the assumed relation $p=w\rho c^2$) as $n=3(1+w)$. A pressure-less, dust filled universe has $w=0$ and hence $n=3$. For a universe filled with relativistic matter, $w=1/3$ and $n=4$. For a universe which contains only vacuum energy with $w=-1$, we have $n=0$ and hence the energy density remains a constant.

   Varying $L$ with respect to $N$, the lapse function, and $a$, the scale factor, we get, respectively, 

\begin{equation}
\frac{\dot{a}^2}{a^2}+\frac{kc^2}{a^2}=\frac{8\pi G}{3}\frac{C_n}{a^n} \label{eq:fried_1}
\end{equation}
and 

\begin{equation}
\frac{2\ddot{a}}{a}+ \frac{\dot{a}^2}{a^2}+\frac{kc^2}{a^2}=\frac{8\pi G}{3}\frac{C_n (3-n)}{a^n},
\end{equation}
which are the relevant classical constraint equations in this case.  The conjugate momentum to $a$ may now be defined as

\begin{equation}
\Pi_a =\frac{\partial L}{\partial \dot{a}} = -\frac{3\pi}{2G} c^2a\dot{a}. \label{eq:Pi}
\end{equation}
The Hamiltonian is

\begin{equation}
H = \Pi_a \dot{a} -L \
=-\frac{G}{3\pi c^2}\frac{\Pi_a^2}{a}- \frac{3\pi }{4G}c^4 ka+2\pi^2  c^2\frac {C_n}{a^{n-3}}. \label{eq:H}
\end{equation}

The WD equation can be written by making the operator replacements for $\Pi_a$ and $a$ in $H$. However, finding the operator corresponding to $\Pi_a^2/a$ is problematic due to an operator ordering ambiguity. The most commonly used form is

\begin{equation}
\frac{\Pi_a^2}{a} \rightarrow -\hbar^2a^{-r-1}\frac{\partial}{\partial a}\left( a^r \frac{\partial}{\partial a} \right)
.\label{eq:op_order}
\end{equation} 
where the choice of $r$ is arbitrary and is usually  made according to convenience.  Using this expression with $r=0$, we obtain the WD equation as

\begin{equation}
\frac{d^2\Psi}{da^2} - \left( \frac{9\pi^2kc^6}{4G^2\hbar^2}a^2 -\frac{6\pi^3c^4}{G\hbar^2} \frac{C_n}{a^{n-4}}\right)\Psi =0.
\end{equation}
The corresponding quantum cosmological Hamilton-Jacobi equation (QCHJE) is obtained by putting $\Psi =\exp({i\hat{S}/\hbar})$ in the above. This gives 

\begin{equation}
\left(\frac{d\hat{S}}{da}\right)^2+ \left( \frac{9\pi^2kc^6}{4G^2}a^2 -\frac{6\pi^3c^4}{G} \frac{C_n}{a^{n-4}}\right)=i\hbar\frac{d^2\hat{S}}{da^2} \label{eq:qchje1}
\end{equation}           
Another frequent choice used in (\ref{eq:op_order}) is $r=-1$, which results in 

\begin{equation}
\frac{d^2\Psi}{da^2} -\frac{1}{a}\frac{d\Psi}{da} - \left( \frac{9\pi^2kc^6}{4G^2\hbar^2}a^2 -\frac{6\pi^3c^4}{G\hbar^2} \frac{C_n}{a^{n-4}}\right)\Psi =0.
\end{equation}
This leads to a QCHJE of the form

\begin{equation}
\left(\frac{d\hat{S}}{da}\right)^2+ \left( \frac{9\pi^2kc^6}{4G^2}a^2 -\frac{6\pi^3c^4}{G} \frac{C_n}{a^{n-4}}\right)=i\hbar \left(\frac{d^2\hat{S}}{da^2}-\frac{1}{a}\frac{d\hat{S}}{da} \right).\label{eq:qchje2}
\end{equation}  

When   equations (\ref{eq:qchje1}) and (\ref{eq:qchje2}) are compared  with the classical Hamilton-Jacobi equation in this case, it is seen that the second terms on their left hand sides  correspond to the classical potential and the terms on the right hand sides correspond to the quantum potential $Q(a)$. Without the latter terms, these equations reduce to the classical Hamilton-Jacobi equation for the Friedmann model. What we look for is a classical potential that can give identical quantum and classical trajectories. It is easily seen that the vanishing of the quantum potential is also a necessary condition for exact classical-quantum correspondence.

We note that both the quantum cosmological  Hamilton-Jacobi equations (\ref{eq:qchje1}) and (\ref{eq:qchje2}) are nonlinear differential equations, with no apparent simple solutions. However,   the latter form of QCHJE turns out to have an exact solution  for the case with $n=2$, such that $\Psi$ is given by

\begin{equation}
\Psi (a)\propto \exp \left( \pm i \frac{m a^2}{2\hbar}\right).\label{eq:Psi_coasting}
\end{equation}
 Here we have defined a constant $m$ as
 
 \begin{equation}
m=\frac{3\pi}{2G} c^2\left(\frac{8\pi G}{3}C_2 -kc^2  \right)^{1/2}. 
\end{equation}
Note that as in the case of plane waves in quantum mechanics for a  single free particle, the wave function of the form (\ref{eq:Psi_coasting}) allows the dBB (where $\Psi  \equiv R\exp({iS/\hbar})$, with $R$, $S$ real) and MdBB (where $\Psi  \equiv \exp({i\hat{S}/\hbar})$) schemes to give identical solutions for $a(t)$  too. That is, one can identify the same function  ${ma^2}/{2}$ as $S$ and $\hat{S}$ respectively in these schemes. This leads to the same  de Broglie-type equation of motion in them, which  can  be obtained from 

\begin{equation}
\Pi_a =  -\frac{3\pi}{2G} c^2a\dot{a} =  \frac{\partial \hat{S}}{\partial a}=\pm ma.
\end{equation}
Here we have used equation (\ref{eq:Pi}). 
Integrating this  gives the evolution of the scale factor  $a$ with time. The solution is 

\begin{equation}
a=\mp \frac{2G}{3\pi c^2} m t =\mp \left(\frac{8\pi G}{3}C_2 -kc^2  \right)^{1/2} t,\label{eq:coasting}
\end{equation}
which  is the well-known 'coasting evolution' in cosmology. This result, obtained from the dBB/MdBB quantum trajectory approaches, agrees with the classical evolution. One  can  see this  by putting $n=2$ in the classical constraint equation (\ref{eq:fried_1}) and integrating for the trajectories. Thus we have exact quantum-classical correspondence for this Friedmann  model with total energy density varying as $1/a^2$.  

We also check the behaviour of the quantum potential $Q(a)$ in this case. It may be noted that for the operator ordering $r=-1$, 

\begin{equation}
Q(a) =- i\hbar \left(\frac{d^2\hat{S}}{da^2}-\frac{1}{a}\frac{d\hat{S}}{da} \right),
\end{equation}
is equal to zero, given the above solution of $\hat{S}=ma^2/2$. This further shows the desired feature of identical classical and quantum evolution for $n=2$.

The solution of the form (\ref{eq:coasting}) commonly appears in the  Milne model \cite{milne}, but since this is an empty model devoid of any matter/energy, much attention was not drawn towards it. The present model has the evolution of scale factor coinciding with that of the Milne model, but it is not an empty one. This is a Friedmann  model with total energy density varying as $1/a^2$. It may be argued that also such  an energy, with equation of state parameter $w=-1/3$, is unrealistic. But it was noted in \cite{jj} that this model can describe the universe when we have two components for the energy density of the universe; one of which is  ordinary matter (with $w= 0$ or $1/3$), and the other is a decaying vacuum energy (with $w=-1$). It was shown that such a model has very close predictions agreeing with recent apparent magnitude-redshift observations of Type Ia supernovae, and is devoid of the conceptual problems that were noted in standard cosmology during the 1980s \cite{jj,mvjvn,mvj4}.

\section{Discussion}

Writing down the WD equation $\hat{H}\Psi=0$ in a mini-superspace is one of the simplest applications of quantum theory in  problems involving gravity.  The dBB and MdBB schemes show that in ordinary quantum mechanics of a single particle, a plane wave solution of the Schrodinger equation corresponds to the motion of a particle that can be considered classical, having a trajectory with constant momentum. This plane wave solution, obtained for a free particle with $V=0$, is used to describe incoming particles in quantum scattering problems. In this paper we attempt to find a potential that provides a solution to the WD equation such that the evolution of the universe can be considered classical.

For this, we resort to trajectory representations and  search for a potential that gives identical classical and quantum evolution for $a(t)$. An equivalent condition for exact quantum-classical correspondence is the vanishing of the quantum potential. We have found that in a Friedmann model where the total energy density varies as $1/a^n$, the exact correspondence happens for $n=2$. In this case, the WD equation has a simple solution of the form $\Psi =\exp(ima^2/2\hbar)$ and both the dBB and MdBB representations give the result that $a\propto t$. We see that for this kind of total energy/matter density, the classical evolution is also $a \propto t$. Therefore this solution  is unique, having identical classical and quantum evolution, as in the case of free particles described by plane waves in ordinary quantum mechanics. The evolution is called 'coasting' in the literature and is usually associated with the classic 'Milne model', which describes an empty universe. However, the present model is not empty; it is a  Friedmann model with $\rho \propto 1/a^2$, capable of describing the expansion of the present universe to a very good approximation. 

The approach in this paper is very similar to that in \cite{he_gao_cai}, where the evolution of the scale factor in a quantum cosmological model is computed using the dBB formalism. The results are significant, for \cite{he_gao_cai} presents a rigorous proof that the universe could be spontaneously and irreversibly created from nothing, and that after-wards it proceeds to an inflationary epoch. It is also shown that the exponential expansion stops when the scale factor grows sufficiently large, no matter whether it is closed, open, or flat. It is of particular interest to note that for the open case in this scenario, after the inflationary epoch, the universe enters a coasting evolution $a\propto t$ and continues to expand in this fashion. The spontaneous creation of the coasting universe obtained in the present paper from nothing is an attractive problem worth pursuing. This special case would have benefits such as no singularity, creation of particles by the decay of vacuum energy, a low dark matter/dark energy budget for the present universe, etc.

Solutions having exact classical-quantum correspondence, as obtained in this paper, are important for they  save us from the conceptual difficulties in applying the quantum theory to the entire universe. 

\noindent {\bf Acknowledgement}

 The author wishes to thank the University Grants Commission, New Delhi for a Research Grant (MRP(S) 1207/11-12/KLMG022/UGC-SWRO).

  \end{document}